\documentstyle[12pt]{article}

\renewcommand{\thefootnote}{\fnsymbol{footnote}}

\begin{document}

%
\newcommand{\text}{\textstyle}
\newcommand{\tprod}{\prod}

\begin{titlepage}
\begin{center}                                                                  
{  \bf  BROWNIAN MOTION ON A SMASH LINE}\renewcommand{\thefootnote}{\dagger} \footnote{Submitted to Journal of Nonlinear Mathematical Physics.
 Special Issue of Proccedings of NEEDS'99.}

\vskip 1.0 cm                                                                   
                                                                                
{\bf Demosthenes Ellinas}\renewcommand{\thefootnote}{*}\footnote{     
Email: {\tt ellinas@science.tuc.gr}} 
and \
{\bf Ioannis Tsohantjis}\renewcommand{\thefootnote}{\natural}
\footnote{     
Email: {\tt ioannis2@otenet.gr}}
\vskip 0.2cm
 Department of Sciences\\
Technical University of Crete \\ 
GR-73 100 Chania Crete Greece\\              
\vskip 0.2cm


\begin{abstract}
Brownian motion  on a smash line algebra (a smash or braided version of the 
algebra resulting by tensoring the  real line and the generalized paragrassmann line algebras), 
is constructed by means of its Hopf algebraic structure. Further, statistical moments, non stationary 
generalizations  and its diffusion
limit are also studied. The ensuing diffusion equation posseses triangular matrix realizations.
\end{abstract}
\end{center}
\end{titlepage}

{\it Introduction.}  The study of random walks and diffusions on generalized
spaces is an active field of research. The formulation of the related
problems is carried out mostly in a algebraic manner that employs commuting
and/or non-commuting algebras with rich structures such as bialgebras and
Hopf algebras with possible braiding or smashing defined among them cf.\cite
{qprobability1,qprobability2,mbook}. Here we construct and study further
such a Brownian motion on a smash line algebra.  Our algebraic model for the
braided/smash line is the merging of the $*$-Hopf algebra  of formal power
series ${\cal A}={\bf C}[[x]]$ on ${\bf R}$, with the braided $*$-Hopf 
algebra ${\cal B}={\bf C}[\xi ]/\xi^N$, of $(N-1)$-degree polynomials
generated by the $N$-potent variable $\xi$ ($N=2$ gives us the standard
supespace). Technically this  merging is the cross-product (also known as
smash-product) of ${\cal A, B}$ algebras.  It imposes a definite braiding
rule between ${\cal A, B}$ that is naturally interpreted  as non
commutativity among the increments (steps) of the underlying random walk. 
Once a natural positive definite functional has been chosen on the smash
line we compute the values of the  statistical moments of any order for our $%
(x, \xi)$ random variables. Introducing an  appropriate limiting procedure
we construct the (pseudo)differential equation of  the diffusion on the
smash line. This is further extended to non stationary random walks that 
incorporate Hamiltonian dynamics into the diffusions processes. 
Additionally a matrix realizations of the $\xi$-generators of the ${\cal B}$
algebra and of their differential are employed to cast  the resulting
diffusion equation into the form of a matrix-valued (ordinary)differential
equation  that gives rise to a system of coupled diffusion equations.


{\it Smash line algebra}. Let us consider two complex associative algebras: $%
{\cal A}={\bf R}[[x]]$, the algebra of real formal power series in one
variable generated by the element $x$ with $\{1,x,x^2,\ldots \}$ as a linear
basis, and ${\cal B}={\bf R}[\xi ]/\xi ^N$ the algebra of polynomials in one
variable of degree $N-1$, generated by the element $\xi $ with a basis $%
\{1,\xi ,{\xi }^2,\ldots {\xi }^{N-1}\}$ and ${\xi }^N=0$. Both these
algebras are equipped with a Hopf algebra structure. Algebras ${\cal A}=%
{\cal A}(\mu _{{\cal A}},\Delta _{{\cal A}},u_{{\cal A}},\varepsilon _{{\cal %
A}})$, ${\cal B}={\cal B}(\mu _{{\cal B}},\Delta _{{\cal B}},u_{{\cal B}%
},\varepsilon _{{\cal B}})$ are commuting, cocommuting coassociative
algebras with products $\mu _{{\cal A}},$ $\mu _{{\cal B}},$, coproducts, $%
\Delta _{{\cal A}},\Delta _{{\cal B}},$ units $u_{{\cal A}},$ $u_{{\cal B}},$%
and counits $\varepsilon _{{\cal A}},$ $\varepsilon _{{\cal B}}$
respectively defined as $\mu _{{\cal A}}(x\otimes y)=xy,$ $\mu _{{\cal B}%
}(\xi \otimes \eta )=\xi \eta ,$ $u_{{\cal A}}(c)=c{\bf 1}_{{\cal A}},$ $u_{%
{\cal B}}(c)=c{\bf 1}_{{\cal B}},$ $c\in {\bf C},$ $\Delta _{{\cal A}%
}(x)=x\otimes {\bf 1}_{{\cal A}}+{\bf 1}_{{\cal A}}\otimes x,$ $\Delta _{%
{\cal B}}(\xi )=\xi \otimes {\bf 1}_{{\cal B}}+{\bf 1}_{{\cal B}}\otimes 
{\bf \xi ,}$ $\epsilon _{{\cal A}}(x)=0,$ $\epsilon _{{\cal A}}({\bf 1}_{%
{\cal A}})=1,\varepsilon _{{\cal B}}(\xi )=0,$and $\varepsilon _{{\cal B}}(%
{\bf 1}_{{\cal B}})=1;$for all $x,y\in {\cal A},$ $\xi ,\eta \in {\cal B}$,
where ${\bf 1}_{{\cal A}},$ ${\bf 1}_{{\cal B}}$ are the identity elements
of ${\cal A}$ and ${\cal B}$ respectively. Moreover a trivial braiding or
twist map $\tau $ is defined in ${\cal A}$ as $\tau (x\otimes y)=y\otimes x$
while in ${\cal B}$ a braiding $\varphi $ ia defined as $\varphi (\xi
\otimes \eta )=q\eta \otimes \xi $ where $q=e^{2\pi iN}$. Random walks
leading to diffusion equations have been carried out independently in ${\cal %
A}$ \cite{maj1}and ${\cal B}$ \cite{maj2,xii}.

We shall now consider the case of performing a random walk on what we shall
call a smash line which arises by first merging of the two algebras above.
The merging of ${\cal A}$ and ${\cal B}$ as $\Omega ={\cal A}\otimes {\cal B}
$ where $x$ is embeded as $x\otimes {\bf 1}_{{\cal A}}$ and $\xi $ as ${\bf 1%
}_{{\cal B}}\otimes \xi $, will eventually be formulated as an associative
smash product algebra $\Omega $ which will also have a smash coproduct
algebra structure. Explicitly the maps of product , coproduct, unit and
counit in $\Omega $ are given by

\begin{equation}
\mu _\Omega \equiv (\mu _{{\cal A}}\otimes \mu _{{\cal B}})\circ (id_{{\cal A
}}\otimes \tau \otimes id_{{\cal B}})\text{ , }\Delta _\Omega \equiv (id_{
{\cal A}}\otimes \tau \otimes id_{{\cal B}})\circ (\Delta _{{\cal A}}\otimes
\Delta _{{\cal B}})  \label{domeg}
\end{equation}

\[
\text{ }u_\Omega (1)={\bf 1}_\Omega \text{ , }\varepsilon _\Omega \equiv
\epsilon _{{\cal A}}\otimes \epsilon _{{\cal B}}\text{ , with }{\bf 1}%
_\Omega ={\bf 1}_{{\cal A}}\otimes {\bf 1}_{{\cal B}} 
\]

We can now give a smash product and coproduct algebra structure\cite
{mbook,mili,drabant}  to $\Omega $ by means of the braiding map $\Psi
:\Omega \otimes \Omega \rightarrow \Omega \otimes \Omega $ where

\[
\Psi (x\otimes {\bf 1}_{{\cal B}}\otimes {\bf 1}_{{\cal A}}\otimes \xi )=Q(
{\bf 1}_{{\cal A}}\otimes \xi \otimes x\otimes {\bf 1}_{{\cal B}}),(Q\in 
{\bf R}), 
\]

\[
\Psi ({\bf 1}_{{\cal A}}\otimes {\bf \xi }\otimes {\bf 1}_{{\cal A}}\otimes
\eta )=q({\bf 1}_{{\cal A}}\otimes \eta \otimes {\bf 1}_{{\cal A}}\otimes 
{\bf \xi }), 
\]

\[
\Psi (x\otimes {\bf 1}_{{\cal B}}\otimes {\it x}^{^{\prime }}\otimes \xi )=(
{\it x}^{^{\prime }}\otimes {\bf 1}_{{\cal B}}\otimes x\otimes {\bf 1}_{
{\cal B}}), 
\]
with product $\mu _{\Omega ^2}=(\mu _\Omega \otimes \mu _\Omega )\circ
(id_\Omega \otimes \Psi \otimes id_\Omega )$, coproduct $\Delta _{\Omega
^2}\equiv (id_\Omega \otimes \Psi \otimes id_\Omega )\circ (\Delta _\Omega
\otimes \Delta _\Omega )$, unit and counit as $u_{\Omega ^2}(1)={\bf 1}
_{\Omega ^2}$ , $\varepsilon _{\Omega ^2}\equiv \epsilon _\Omega \otimes
\epsilon _\Omega $ . This product extended to $\Omega ^n$ by the relation

\[
\mu _{\Omega ^n}=(\mu _{\Omega ^{n-1}}\otimes \mu _\Omega )\circ
\tprod_{k=1}^{2n-3}\circ (id_\Omega ^{\otimes ^{2n-2-k}}\otimes \Psi \otimes
id_\Omega ^k). 
\]
provides the algebra of increments of the random walk on the smash line i.e

\begin{equation}
x_ix_j=x_jx_i\text{, }\forall i,j\text{, }\xi _i\xi _j=q\xi _j\xi _i\text{, }
x_i\xi _j=Q\xi _jx_i\text{, for }i>j  \label{rel}
\end{equation}
where the indices above indicate the position of the embeddings of $x$ and $
\xi $ in the respective spaces (e.g. $\xi _2\xi _1=({\bf 1}_{{\cal A}
}\otimes {\bf 1}_{{\cal B}}\otimes {\bf 1}_{{\cal A}}\otimes \xi )({\bf 1}_{
{\cal A}}\otimes {\bf \xi }\otimes {\bf 1}_{{\cal A}}\otimes {\bf 1}_{{\cal B
}})$

$=q({\bf 1}_{{\cal A}}\otimes {\bf \xi }\otimes {\bf 1}_{{\cal A}}\otimes 
{\bf 1}_{{\cal B}})({\bf 1}_{{\cal A}}\otimes {\bf 1}_{{\cal B}}\otimes {\bf 
1}_{{\cal A}}\otimes \xi )=q\xi _1\xi _2$). In addition using eqs. (\ref
{domeg}) and (\ref{rel}), the $n$-th fold coproduct on $x^k\otimes {\xi }^l$ 
$\in $ $\Omega $ is given by: 
\[
\Delta _\Omega ^{n-1}(x^k\otimes {\xi }^l)=\sum_{i_1+\cdots +i_n=k}\text{ }
\sum_{j_1+\cdots +j_n=k}\text{ }\left( 
\begin{array}{ccc}
& k &  \\ 
i_n & \cdots & i_n
\end{array}
\right) \left[ 
\begin{array}{ccc}
& l &  \\ 
j_n & \cdots & j_n
\end{array}
\right] _q 
\]

\begin{equation}
\times x^{i_1}\otimes {\xi }^{j_1}\otimes \cdots \otimes x^{i_n}\otimes {\xi 
}^{j_n}  \label{ndelta}
\end{equation}
where ingeneral $\Delta ^n=(\Delta \otimes id^{\otimes ^{n-1}})\circ \Delta
^{n-1}=(id^{\otimes ^{n-1}}\otimes \Delta )\circ \Delta ^{n-1}$, and the $q$
-binomial coefficient is defined as $\left[ 
\begin{array}{l}
m \\ 
kl
\end{array}
\right] _q=\frac{[m]_q!}{[k]_q![l]_q!}$.

{\it Diffusion equation}. Let $\phi $ be a linear functional from our
algebra to ${\bf C}$, which corresponds to probability denisty element $\rho 
$ satisfing the following relations:

\[
\phi (f)=<f>_\phi =\int \rho f=<\phi \text{ , }f>\text{ }\in {\bf C}\text{.} 
\]
{It is assumed that }$\phi $ lives in the dual space of $\Omega $ where the
product is defined to be the usual convolution operation between probability
density functions, by means of the following relations: 
\[
\phi ^{*n}(f)=<\phi ^{\otimes ^n},\Delta ^{n-1}(f)> 
\]
where $\Delta ^n\equiv \Delta _\Omega ^n$ . If this is interpreted as the
state of probability function of the random walk after $n$ steps, then the
general state after an n-step walk evaluated on a general observable element

\begin{equation}
f(x,\xi )=\sum_{k,l\in N}d_{kl}x^k\xi ^l+\sum_{i\in Z_{+}}b_ix^i+\sum_{j\in
Z_{+}}c_j\xi ^j  \label{f}
\end{equation}
of $\Omega $ reads,

\begin{equation}
\phi ^{*n}(\text{ }f(x,\xi ))=(\phi *\phi *...*\phi )(\text{ }f(x,\xi ))=({
\phi }^{\otimes ^n})\Delta ^{n-1}(f(x,\xi )).  \label{fi}
\end{equation}
{By virtue of eq.(\ref{ndelta}), (\ref{fi}) and making the choice $\rho
(x,\xi )=\rho _1(x)\rho _2(\xi )$ where ${\rho }_1(x)=p_1\delta
(x-a)+(1-p_1)\delta (x+a)$, ${\rho }_2(\xi )=p_2\delta (\xi -\theta
)+(1-p_2)\delta (\xi +\theta )$ for the density functions of the random walk
where as usual }$p_1$ and $p_2$ are chosen probablilities, {we conclude that
with respect to the $x\xi $-monomials 
\[
<x^k\otimes {\xi }^l>_{{\phi }^{*n}}=\sum_{i_1+\cdots +i_n=k}\text{ }
\sum_{j_1+\cdots +j_n=k}\text{ }\left( 
\begin{array}{ccc}
& k &  \\ 
i_n & \cdots & i_n
\end{array}
\right) \left[ 
\begin{array}{ccc}
& l &  \\ 
j_n & \cdots & j_n
\end{array}
\right] _q 
\]
}

\begin{equation}
\times \tprod_{l=1}^n<x^{i_l}\otimes {\xi }^{j_l}>_\phi  \label{xiksi}
\end{equation}
{where in general 
\[
<x^i\otimes {\xi }^j>_\phi = 
\]
\[
\left[ p_1e^{aD_x}+(1-p_1)e^{-aD_x}\right] \left[ p_2e_q^{\theta D_\xi
}+(1-p_2)e_q^{-\theta D_\xi }\right] |_{x,\xi =0}(x^i{\xi }^j)={\phi }_x(x^i)
{\phi }_\xi ({\xi }^j)\;. 
\]
}

{Let us now compute the system after }${n}${\ steps and its limit }as $
n\rightarrow \infty $ . {Using Taylors expansion, the form }${{\phi }^{*n}}$ 
{for the case of monomials of }${x^m}${, }${{\xi }^t}$ {and }${x^k{\xi }^l}$ 
{reads as follows: 
\begin{eqnarray}
{\phi }^{*n} &=&{\phi }_x^{*n}{\phi }_\xi ^{*n}=[\varepsilon _\Omega
+2a(p_1-1/2)D_x+a^2/2!D_x^2+...]^n  \nonumber \\
&&\times [{\varepsilon }_\Omega +2\theta (p_2-1/2)D_\xi +{\theta }
^2/[2]_q!D_\xi ^2+...]^n|_{x,\xi =0}  \label{phin}
\end{eqnarray}
and where $D_x=\partial /\partial _x,$ $D_\xi f(\xi )=\frac{f(\xi )-f(\xi q)
}{(1-q)\xi }$. Following \cite{maj1}, \cite{maj2} we substitute $2a(p_1-1/2)=
\frac{c_1t}n$, $2{\theta }(p_2-1/2)=\frac{c_2t}n$, $a^2 /2=\frac{{\alpha }_1t
}n$, $\theta^2/[2]_q=\frac{{\alpha }_2t}n$ and then we take the limit $
n\rightarrow \infty $ with $t$, $c_1$, $c_2$, ${\alpha }_1$, ${\alpha }_2$
fixed and $t=n\delta $, $\delta $ being the size of the step in time, to
obtain the continue limit of random walk where the steps are viewed (\cite
{maj1}, \cite{maj2}) as steps in time. This yields \ 
\begin{equation}
{\phi }^\infty (f)=(e^{c_1tD_x+{\alpha }_1tD_x^2}f)\mid _{x=0}+(e^{c_2tD_\xi
+{\alpha }_2tD_\xi ^2}f)\mid _{\xi =0}+(e^{c_1tD_x+{\alpha }
_1tD_x^2+c_2tD_\xi +{\alpha }_2tD_\xi ^2}f)\mid _{x,\xi =0}  \nonumber
\end{equation}
where the limit $(1+z/n)^n\rightarrow e^z$ for $n\rightarrow \infty$ has
been used. }To obtain the diffusion equation we have, that for a general $f$
of the form {(\ref{f}),\ 
\begin{eqnarray}
\int (\partial _t\rho ^\infty )f &=&\partial _t\phi ^\infty (f)=\phi ^\infty
((c_1D_x+{\alpha }_1D_x^2+c_2D_\xi +{\alpha }_2D_\xi ^2)f)  \nonumber \\
&=&\int ((-c_1D_x+{\alpha }_1D_x^2+c_2D_\xi ^{*}+{\alpha }_2D_\xi ^{*2})\rho
^\infty )f
\end{eqnarray}
which implies that 
\begin{equation}
\partial _t\rho ^\infty =(-c_1D_x+{\alpha }_1D_x^2+c_2D_\xi ^{*}+{\alpha }
_2D_\xi ^{*2})\rho ^\infty  \label{dif1}
\end{equation}
}with solution:{\ } 
\begin{equation}
\rho _1^\infty (a)=(4{\pi }{\alpha }_1t)^{-1}e^{-\frac{(a-c_1t)^2}{4{\alpha }
_1t}}\text{ , }\rho _2^\infty (\theta )=\sum_{k=0}^{N-1}{\theta }
^{N-1-k}\sum_{l=0}^{l<k/2}\frac{(c_2t)^{k-l}({\alpha }_2/c_2)^l[k]_q!}{
l!(k-2l)!}
\end{equation}

{\it Non Stationary Case}. Consider the Hamiltonian evolution of a quantity
F depending in general on the phase space variables $x$, $p$, $\xi $, $p_\xi 
$ where we have assumed that $\xi ^2=0$ :

\begin{equation}
\{F,H\}=\frac{\partial F}{\partial x}\frac{\partial H}{\partial p}-\frac{
\partial F}{\partial p}\frac{\partial H}{\partial x}+(-1)^\epsilon (\frac{
\partial F}{\partial \xi }\frac{\partial H}{\partial p_\xi }+\frac{\partial F
}{\partial p_\xi }\frac{\partial H}{\partial \xi })
\end{equation}
where $\epsilon =0,1$ is the degree of $F$. We can defined, using $t$ as a
time parameter, the action of $e^{tV_H}$ on $x_0$ is given by $
e^{tV_H}x_0=x_t$, and similarly $e^{tV_H}\xi _0=\xi _t$ where the action of $
V_H$ isgenerally given by

\[
V_H(F)=\frac{\partial H}{\partial p}\frac{\partial F}{\partial x}-\frac{
\partial H}{\partial x}\frac{\partial F}{\partial p}+(-1)^\epsilon (\frac{
\partial H}{\partial p_\xi }\frac{\partial F}{\partial \xi }-\frac{\partial H
}{\partial \xi }\frac{\partial F}{\partial p_\xi }) 
\]

We can now evaluate $\phi (x_t^k\xi _t^l)$ using the choice of $\rho $ as
mentioned

underneath relation ({\ref{fi})}:

\begin{equation}
\phi (x_t^k\xi _t^l)=[p_1e^{a_tD_x}+(1-p_1)e^{a_t^{\prime
}D_x}][p_2e^{\theta _\tau d_\xi }+(1-p_2)e^{\theta _\tau ^{\prime }d_\xi
}]_{\mid x,\xi =0}x_0^k\xi _0^l=\phi _t(x_0^k\xi _0^l)
\end{equation}
where we have set $a_t=a-t\lambda $ , $a_t^{\prime }=-a-t\lambda $, $\theta
_t=\theta -t\tilde{\lambda}$ and $\theta _t^{\prime }=-\theta -t\tilde{
\lambda}$, $\lambda =\frac{\partial H}{\partial p}$, $\tilde{\lambda}=-\frac{
\partial H}{\partial p_\xi }$. We can now expand the exponentials in the
above relation and after some algebra which involves taking the limit $
n\rightarrow \infty $ we find that

\begin{equation}
{\phi }_t^\infty (f)=[e^{(c_1-\lambda d_1)tD_x+{\alpha }_1tD_x^2+(c_2-\tilde{
\lambda}d_2)tD_\xi }](f)_{\mid x,\xi =0}=(e^{tK}f)_{\mid x,\xi =0}=\int \rho
^\infty f=<\rho ^\infty ,f>
\end{equation}
where we have implemented the following substitutions $2a(p_1-1/2)=\frac{c_1t
}n$, $t\lambda =\lambda \frac{d_1t}n$, $\frac{a^2}2=\frac{\alpha _1t}n$, $
2\theta (p_2-1/2)=\frac{c_2t}n$, $t\tilde{\lambda}=\tilde{\lambda}\frac{d_2t}
n$, ( $d_1$, $d_2$, real contsants).

Then the diffusion equation reads: 
\begin{equation}
\partial _t\rho ^\infty =(-c_1D_x+{\alpha }_1D_x^2+c_2D_\xi ^{*}+\lambda D_x-
\tilde{\lambda}D_\xi ^{*})\rho ^\infty  \label{dif2}
\end{equation}
where we have set $d_1=d_2=1$.

{\it Matrix Realisation}. We will now employ the $N$ -dimentional matrix
representations of $\xi $, as $(\xi )_{i,i+1}=1$, $D_\xi $ as $(D_\xi
)_{i,i+1}=\{i\}$ and $D_\xi ^{*}$ as $(D_\xi ^{*})_{i,i+1}=\{i\}e^{\omega
^{-1}\{i\}}$, $i=1,\cdots ,N-1$ and all other entries zero and where $\{x\}=
\frac{1-\omega ^x}{1-\omega }$, $\omega =e^{2i\pi /N}$, that they have
explicitely been constructed in \cite{rausch,isaev}, to get a matrix
realisaton for the stationary diffusion equation ({\ref{dif1}).} In
particular for the matrix representation of $D_\xi ^{*}(t)$ one should take
in to account \cite{maj2} where the above has been obtained via the use of
anyonic Leibnitz rule $D_\xi (fg)=(D_\xi f)g+L_qf(D_\xi g)$ and the property
that $D_\xi ^{*}=-D_\xi L_{q^{-1}}$ .Using this realization we can write (
\ref{dif1}) as ${\partial _t\rho ^\infty =H}\rho ^\infty $ , where

\[
H=\left( 
\begin{array}{lllll}
H_x & c_2\lambda _1 & \alpha _2\lambda _1\lambda _2 & 0 & 0 \\ 
0 & H_x & c_2\lambda _2 & \alpha _2\lambda _2\lambda _3 & 0 \\ 
\vdots & \vdots & \vdots & \vdots & \vdots \\ 
0 & 0 & 0 & H_x & c_2\lambda _{N-1} \\ 
0 & 0 & 0 & 0 & H_x
\end{array}
\right) , 
\]
with $\lambda _i=\{i\}e^{\omega ^{-1}\{i\}}$ , $i=1,\cdots ,N-1$, and $
H_x=-c_1D_x+{\alpha }_1D_x^2$ . Expanding a generic $\rho ^\infty $ as:

\[
\rho ^\infty (x,\xi ,t)=\sum_{i=0}^{N-1}\sum_{j=0}^\infty \rho _{ij}x^j\xi
^i=\sum_{i=0}^{N-1}\rho _i(x)\xi ^i=\left( 
\begin{array}{llll}
\rho _0(x) & 0 & 0 & 0 \\ 
\rho _1(x) & \rho _0(x) & 0 & 0 \\ 
\vdots & \vdots & \vdots & \vdots \\ 
\rho _{N-1}(x) & \rho _{N-2}(x) & \ldots & \rho _0(x)
\end{array}
\right) , 
\]
yields the following general system of differential equations to be solved:

\begin{equation}
\frac{\partial \rho _k}{\partial t}=H_x\rho _k+c_2\lambda _{k+1}\rho
_{k+1}+\alpha _2\lambda _{k+1}\lambda _{k+2}\rho _{k+2} \ \ \ \ {\rm for \ \
\ k=0,1,\cdots ,N-1 \;.}  \label{system}
\end{equation}

{\it Conclusions}. We have constructed an algebraic random walk and its
associated limit governed by a diffusion equation on a space with real and
paragrassmann components. The construction is based on the smashing of the
algebra of functions of the underlined space. The algebraic approach is
flexible and allows to determine statistical moments of the random walk and
matrix realizations of its diffusion limit. Details of the smash line
Brownian motion as well as extensions to random walks and diffusions on
operator algebras of Quantum Mechanics can be found elsewere\cite{elltso,ell}
.

\end{document}